\documentclass[useAMS,usenatbib,usegraphicx,usedcolumn]{mn2e}
\usepackage{fix2col}

\newcommand{\CIV}{\mbox{C\,{\sc iv}}}

\newcommand{\SiIV}{\mbox{Si\,{\sc iv}}}

\newcommand{\MgII}{\mbox{Mg\,{\sc ii}}}

\newcommand{\HeII}{\mbox{He\,{\sc ii}}}

\newcommand{\kms}{km~s$^{-1}$}

\newcommand{\ergs}{erg~s$^{-1}$}

\newcommand{\LLedd}{\mbox{$L/L_{\rm Edd}$}}
\newcommand{\fl}{\mbox{$f_\lambda$}}
\newcommand{\Mbh}{\mbox{$M_{\rm BH}$}}
\newcommand{\auv}{\mbox{$\alpha_{\rm UV}$}}
\newcommand{\aox}{\mbox{$\alpha_{\rm ox}$}}

\newcommand{\vs}{\mbox{$v_{\rm shift}$}}
\newcommand{\vsp}{\mbox{$v_{\rm md}$}}

\newcommand{\lamrest}{\mbox{$\lambda_{\rm rest}$}}
\newcommand{\fnrepeat}[1]{$^{\ref{#1}}$}

\newcommand{\rS}{\mbox{$r_{\rm S}$}}

\title[On the origins of \CIV\ BAL diversity]{On the origins of \CIV\ absorption profile diversity in broad absorption line quasars}

\author[A.~Baskin, A.~Laor and F.~Hamann]
{Alexei Baskin,$^1$\thanks{E-mail: alexei@physics.technion.ac.il} 
Ari Laor$^1$ and Fred Hamann$^2$ \\
$^1$Physics Department, Technion -- Israel Institute of Technology, Haifa~32000, Israel\\
$^2$Department of Astronomy, University of Florida, Gainesville, FL 32611-2055, USA}

\begin{document}
\date{}
\pagerange{\pageref{firstpage}--\pageref{lastpage}} \pubyear{2015}
\maketitle
\label{firstpage}

\begin{abstract}
There is a large diversity in the \CIV\ broad absorption line (BAL) profile among BAL quasars (BALQs). We quantify this diversity by exploring the distribution of the \CIV\ BAL properties, FWHM, maximum depth of absorption and its velocity shift (\vsp), using the SDSS DR7 quasar catalogue. We find the following: (i)~Although the median \CIV\ BAL profile in the quasar rest-frame becomes broader and shallower as the UV continuum slope (\auv\ at 1700--3000~\AA) gets bluer, the median individual profile in the absorber rest-frame remains identical, and is narrow (${\rm FWHM}=3500$~\kms) and deep. Only 4 per cent of BALs have ${\rm FWHM}>10,000$~\kms. (ii)~As the \HeII\ emission equivalent-width (EW) decreases, the distributions of FWHM and \vsp\ extend to larger values, and the median maximum depth increases. These trends are consistent with theoretical models in which softer ionizing continua reduce overionization, and allow radiative acceleration of faster BAL outflows. (iii)~As \auv\ becomes bluer, the distribution of \vsp\ extends to larger values. This trend may imply faster outflows at higher latitudes above the accretion disc plane. (iv)~For non-BALQs, the  \CIV\ emission line decreases with decreasing \HeII\ EW, and becomes more asymmetric and blueshifted. This suggests an increasing relative contribution of emission from the BAL outflow to the \CIV\ emission line as the ionizing spectral energy distribution (SED) gets softer, which is consistent with the increasing fraction of BALQs as the ionizing SED gets softer.

\end{abstract}

\section{Introduction}\label{sec:intro}
Broad absorption line quasars (BALQs) are a subtype of quasars, characterized by the presence of broad and blueshifted absorption features \citep*{weymann_etal81}. The intrinsic fraction of quasars that are BALQs is estimated to be typically $\sim$20 per cent (\citealt{hewett_foltz03, reichard_etal03, knigge_etal08, gibson_etal09}; cf.\ \citealt{allen_etal11}), and ranges from  $\sim$4 per cent in blue, strong \HeII\ emission quasars, to $\sim$30 per cent in red, weak \HeII\ emission quasars \citep*[hereafter BLH13]{baskin_etal13}. Although there are several differences in emission properties between BALQs and non-BALQs, the two subtypes appear to be drawn from the same population \citep*{weymann_etal91, hamann_etal93, reichard_etal03}. The \CIV\ broad absorption line (BAL) spans a large range in depth, width and the velocity shift (\vs) of maximum absorption depth  (\vsp) among different BALQs. 

What produces the large diversity of \CIV\ BAL properties? The maximal outflow velocity of \CIV\ BAL has a clear correlation with the quasar luminosity in X-ray weak objects \citep*{brandt_etal00, laor_brandt02, gibson_etal09}. The \CIV\ BAL absorption is on average stronger in BALQs that also show lower ionization BALs \citep{reichard_etal03, filizAk_etal14}. Recently, BLH13 studied the median absorption properties of BALQs, and found that the width and \vs\ of \CIV\ BAL are set by the \HeII\ $\lambda$1640 emission equivalent-width (EW), while the BAL depth is controlled by the UV continuum slope in the 1700--3000~\AA\ range (\auv). The explored BAL profiles in BLH13 are \emph{median} profiles, derived from composite spectra of BALQs from the Sloan Digital Sky Survey (SDSS; \citealt{york_etal00}) Data Release 7 (DR7; \citealt{abazajian_etal09, schneider_etal10, shen_etal11}).

In this paper, we extend the study of BLH13, and explore the dependence of the \CIV\ BAL profile on the \HeII\ EW and \auv\ for \emph{individual} absorption features in BALQs from the same SDSS DR7 sample ($1.75\leq z \leq 2.05$). We find that in contrast with the median \CIV\ BAL profiles which have ${\rm FWHM}>5000$~\kms, the individual profiles are rather narrow with a median ${\rm FWHM}=3500$~\kms. We also find that \auv\ has a significantly weaker effect on the BAL depth, and that the \HeII\ EW has a weaker effect on the BAL FWHM, compared to the effects implied by the median profiles in BLH13. As we show below, these differences result from the strong dependence of \vsp\ on both \auv\ and the \HeII\ EW.

The motivation to use the \HeII\ EW and \auv\ is driven by both observational and theoretical considerations (see BLH13 for a more extensive discussion). Observationally, BALQs are reported to have a weaker \HeII\ EW \citep{richards_etal02, reichard_etal03} and a redder \auv\ compared to non-BALQs \citep{reichard_etal03, maddox_etal08, gibson_etal09, allen_etal11}. Theoretically, the EW of the \HeII\ $\lambda$1640 recombination line is a measure of the continuum strength above 54~eV, compared to the near-UV continuum, and thus is indicative of the hardness of the ionizing spectral energy distribution (SED). The UV slope is reported to correlate with other reddening indicators (e.g.\ \citealt{baskin_laor05, stern_laor12}), and is potentially indicative of our viewing angle, with objects observed closer to edge-on having a redder \auv\ (BLH13). BLH13 show that a Small Magellanic Cloud extinction law with $A_V=0.06$~mag can explain the median reddening of BALQs compared to non-BALQs in a large range of $\lamrest=1000-3000$~\AA, further supporting the interpretation of \auv\ as a dust-reddening indicator.

The paper is structured as follows: In Section~\ref{sec:data_analysis}, we describe the data analysis method. The results are presented in Section~\ref{sec:results} and discussed in Section~\ref{sec:discussion}. Section~\ref{sec:conclusions} summarizes our conclusions.

\section{The Data Analysis}\label{sec:data_analysis}

A complete description of the sample and the data analysis is provided in BLH13, and is briefly reviewed here. The data set is drawn from the SDSS DR7 quasar catalogue of \citet{shen_etal11}. In this catalogue, the BALQ classification is based on the \citet{gibson_etal09} BALQ catalogue for objects that are included in the SDSS DR5, and on a visual inspection of the \CIV\ region for the remaining $\simeq$20 per cent of objects. An object is flagged as a BALQ by \citet{gibson_etal09}, if it has ${\rm BI}_0>0$, where ${\rm BI}_0$ is a modified version of `balnicity index' (BI; \citealt{weymann_etal91}). For ${\rm BI}_0$, the integration over the normalized \fl\ starts at $\vs=0$, rather than at $-3000$~\kms\ which is used for BI. We require the spectra to be in the $1400\leq \lamrest \leq 3000$~\AA\ range, i.e. $1.75\leq z \leq 2.05$ for the SDSS, which allows to cover the \CIV\ BAL and $\lamrest=3000$~\AA\ that is used for \auv, and to exclude LoBALQs using \MgII. This criterion produces a sample of 1691 BALQs and 13,388 non-BALQs. We further require the objects to have a signal-to-noise ratio S/N$\geq$3 in the SDSS {\it i}-filter, to avoid unusually low-S/N spectra, which excludes 39 BALQs and 739 non-BALQs. Finally, we exclude 56 LoBALQs with a \MgII\ BAL detection \citep{shen_etal11}, and construct a sample of HiBALQs only. The final data set is comprised of 1596 HiBALQs and 12,649 non-BALQs.

It has been suggested that radio-loud (RL) BALQs have different BAL properties than radio-quite (RQ) BALQs (e.g.\ \citealt{becker_etal97, brotherton_etal98}). In contrast, \citet{rochais_etal14} find that RL and RQ BALQs are not fundamentally different objects. Here, we use the radio loudness parameter $R\equiv f_{\rm 6\,cm}/f_{2500}$ provided in the \citet{shen_etal11} catalogue, where $f_{\rm 6\,cm}$ is estimated from the FIRST integrated flux density at 20~cm \citep*{becker_etal95, white_etal97}, assuming $f_\nu\propto\nu^{-0.5}$. Objects with $R>10$ are classified as RL (89 BALQs and 782 non-BALQs in our sample). The fraction of RL objects is $\la10$ per cent in the BALQ and non-BALQ subsamples explored here, and thus has a little effect on our results. In the subsequent analysis, we mark RL BALQs, but do not exclude them from our sample. As we show below, RL BALQs have a preference toward a smaller \HeII\ EW and a redder \auv\ compared to RQ BALQs, which results in a different `typical' BAL profile. However, when RL and RQ BALQs are matched in the emission parameters (see also \citealt{rochais_etal14}, where the objects are matched in $L$), there are no systematic differences in the \CIV\ BAL profile between the two subclasses. This implies that radio loudness has no direct effect on the BAL properties. 

The following procedure is carried out for each object (both a BALQ and a non-BALQ): 
\begin{enumerate}
 \item The spectrum is smoothed by a 22 pixel-wide moving average filter, which is equivalent to $\simeq$1350~\kms\ (i.e.\ $\sim$9 resolution elements; \citealt{york_etal00}). This relatively broad filter smooths out any narrow features ($\la$500~\kms) superimposed on the \CIV\ BAL, and diminishes their effect on the measurement of maximum depth of the \CIV\ BAL (see below).
 \item The spectrum is normalized by the mean \fl\ in the $\lamrest=1700-1720$~\AA\ range. 
 \item The \HeII\ emission EW is measured by integrating the normalized \fl\ in the $\lamrest = 1620-1650$~\AA\ range, and approximating $\fl^{\rm cont}$ by a constant value of 1 between the normalization window and 1620~\AA. This rough approximation of $\fl^{\rm cont}$ does not affect significantly our results (see BLH13).
 \item The UV spectral slope \auv\ ($f_\nu \propto \nu^{\alpha_{\rm UV}}$) is measured between the 1700--1720 and 2990--3010~\AA\ windows, using the mean \fl\ of each window. The accuracy of \auv\ is affected by the flux calibration of individual spectra. The flux is calibrated by matching the spectra of simultaneously observed standard stars to the magnitude of their point spread function, and it is accurate to a level of a few per cent for $\lambda_{\rm obs}>4000$~\AA\ \citep{Adelman_McCarthy_etal07, Adelman_McCarthy_etal08}. For $\lambda_{\rm obs}<4000$~\AA, \citet{paris_etal11} report systematic excess light in the SDSS DR7 spectra (see also \citealt{abazajian_etal09}). Since $\lambda_{\rm obs}=4000$~\AA\ corresponds to $\lamrest=1450$~\AA\ for the lowest $z=1.75$ of our sample, the measurement of \auv\ is not affected by this systematic excess. 
\end{enumerate}
We take from the \citet{shen_etal11} catalogue the values of $L(3000$~\AA), \MgII\ FWHM and \Mbh, which is evaluated using the \citet{vestergaard_osmer09} prescription. We also calculate $\log\LLedd = \log L(3000\mbox{~\AA})-\log\Mbh-37.4$, where we use a bolometric correction factor of 5.15 from \citet{shen_etal08}.

\begin{figure}
 \includegraphics[width=84mm]{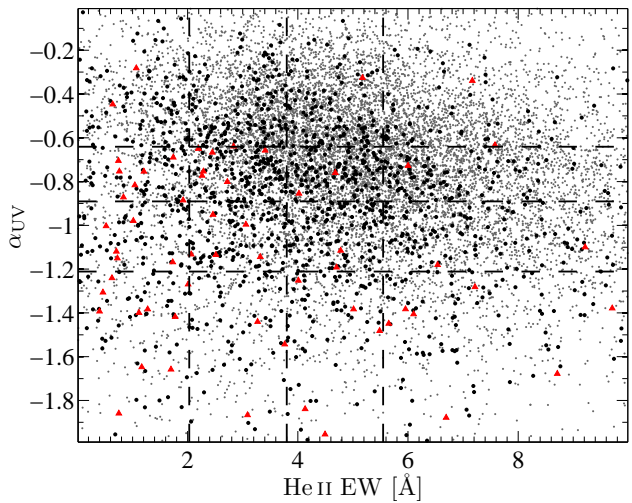}
\caption{
The distribution of \auv\ versus the \HeII\ EW for the sample of BALQs (large black dots for RQ and red triangles for RL) and non-BALQs (small grey dots) which are used in this study. The dashed lines are the boundaries that define the four \HeII\ EW bins and the four \auv\ bins.
}\label{fig:binning}
\end{figure}

The objects are binned based on the \HeII\ EW and \auv, into four bins for each parameter. The binning is such that each BALQ bin contains the same number of objects. The matching non-BALQ bins cover the same parameter range, but do not have equal number of objects per bin. Fig.~\ref{fig:binning} presents the distribution of \auv\ versus the \HeII\ EW for the whole sample of BALQs and non-BALQs, and the boundaries that are used to define the bins. The \HeII\ EW and \auv\ distributions for BALQs and non-BALQs within each bin can be somewhat different, but the difference in the median value of the parameter that is used for binning is below 10 per cent within each bin for all bins (Tables~\ref{tab:HeII_bins} and \ref{tab:auv_bins}). There is no correlation between the \HeII\ EW and \auv\ in BALQs, and a very weak correlation in non-BALQs. The Spearman rank-order correlation coefficient is $\rS=-0.007$ and $-0.139$ for BALQs and non-BALQs, respectively. Note that RL BALQs (Fig.~\ref{fig:binning}, red triangles) have a preference toward lower values of \HeII\ EW and a red \auv.

We measure the maximum absorption depth (CF), velocity shift of the maximum absorption depth (\vsp) and the FWHM for the deepest \CIV\ BAL trough of each BALQ. For brevity, we denote the maximum absorption depth as CF, noting that this notation is formally correct (i.e.\ implies the covering factor) only if $\tau(\vsp)\gg1$. The \vsp\ is measured directly for each BALQ spectrum, and equals the value of \vs\ where $f_\lambda(\lamrest)$ is minimal, with the requirement that $0<\vs<30,000$~\kms. The value of \vs\ is calculated relative to $\lambda=1549$~\AA. The value of CF and the FWHM is estimated as follows:
\begin{enumerate}
 \item A non-BALQ composite is calculated for all the non-BALQ bins (based on the \HeII\ EW and \auv).\footnote{We use a modified median method to calculate the composites. In this method, the adopted value for the composite $f_\lambda$ at a given \lamrest\ equals to the mean $f_\lambda$ of 10 per cent of the objects above and below the median (i.e.\ 20 per cent of objects in total; see BLH13 for details).} This composite is assumed to represent the intrinsically unabsorbed emission for the BALQs which reside in that bin.
 \item Each BALQ spectrum is divided by the non-BALQ composite.
 \item A power law is fit between 1600 and 1800~\AA, and is extrapolated to $\lamrest<1600$~\AA\ (using the 1700--3000~\AA\ range for the fit produces similar results).
 \item The ratio spectrum [from step (ii)] is divided by the power law [from step (iii)], and the result is adopted as the emission-corrected absorption spectrum.
 \item The absorption depth at \vsp\ is adopted as the CF.
 \item Starting from \vsp, we search for the closest $\vs>\vsp$ and $\vs<\vsp$, where the absorption depth is half of that adopted in step (v). The FWHM is the difference between these two values of \vs.
\end{enumerate}
Measuring \vsp\ from the emission-corrected absorption spectrum, i.e.\ after steps (i)--(iv), rather than from the observed spectrum, does not change qualitatively the resulting trends which are explored below. The assumption in step (i) produces $\simeq$10 BALQs per bin with CF smaller than 0.1, which corresponds to ${\rm BI}=0$ (i.e.\ formally not a BALQ). However, we retain these BALQs in our BALQ sample, since reclassifying the objects of the SDSS DR7 quasar catalogue of \citet{shen_etal11} is beyond the scope of our paper. Excluding these BALQs from the BALQ sample does not affect our results. 

As shown in BLH13, the selection by the \HeII\ EW provides a well-matched \CIV\ emission profile on the red side for BALQs and non-BALQs. The selection by \auv\ produces a BALQ/non-BALQ ratio spectrum with a weak pseudo-absorption at a level of $\la$0.05 for $\vs\ga0$. A mismatch in the \CIV\ emission profile has a very small effect on the measurement of BALs with $\vsp>10^4$~\kms, since the \CIV\ emission is generally found at $\vs<10^4$~\kms. As we show below, the BALs with $\vsp < 10^4$~\kms\ tend to have a ${\rm CF} >0.5$ that is well above 0.05. Thus, we neglect the effect of pseudo-absorption. We cannot test whether there are significant differences on the blue side, which will affect the emission-corrected absorption spectrum.

Table~\ref{tab:ind_prop} lists the measured \HeII\ EW, \auv\ and the \CIV\ BAL parameters for individual objects.\footnote{The full version of Table~\ref{tab:ind_prop} is available online in a machine readable format.} Specifically, column (1) lists the SDSS DR7 designation (J2000.0). Columns (2) and (3) provide the right ascension and declination angles (in decimal degrees; J2000.0). Column (4) lists the values of $z$ from \citet{hewett_wild10} which are used in this study. Columns (5) and (6) list the measured values of \HeII\ EW (in \AA) and \auv. Column (7) provides \vsp\ of \CIV\ BAL in units of \kms. Columns (8) and (9) list the \CIV\ BAL CF and FWHM (in \kms) for the \HeII-EW binning. Binning by \auv\ procedures slightly different \CIV\ BAL CF and FWHM for some objects (the measurement of \vsp\ is independent of the binning; see above). However, there is a very good overall agreement between the two binning procedures. Comparing the values of CF from the two procedures yields a Pearson linear-correlation coefficient of $r_{\rm P}=0.997$. A similar analysis for the FWHM produces $r_{\rm P}=0.961$.

\begin{table*}
\begin{minipage}{155mm}
\caption{Measured emission and absorption properties.\fnrepeat{fn1:full}}\label{tab:ind_prop}
\begin{tabular}{@{}{l}*{7}{c}{c}@{}}
\hline
\multicolumn{1}{c}{SDSS J}& R.A. & Dec. & $z$ & \HeII\ EW & \auv\fnrepeat{fn0:auv} & \vsp & CF\fnrepeat{fn0:binning} & FWHM\fnrepeat{fn0:binning}\\
 & (deg) & (deg) & & (\AA) & &  (\kms) & & (\kms) \\
\multicolumn{1}{c}{(1)} & (2) & (3) & (4) & (5) & (6) & (7) & (8) & (9) \\
\hline
000013.80$-$005446.8 & 0.057506 & $-$0.91300 & 1.8409 & 6.8 & $-1.00$ & 2277 & 0.73 & 1932 \\ 
000038.66+011426.2 & 0.161080 &  1.24060 & 1.8330 & 6.1 & $-1.39$ & 7107 & 0.93 & 4899 \\ 
000119.64+154828.8 & 0.331840 & 15.80800 & 1.9211 &  3.3 & $-1.18$ & 8418 & 0.78 & 10695 \\ 
000645.98$-$004840.1 & 1.691600 &  $-$0.81116 & 2.0036 & 7.8 & $-0.93$ & 16905 & 0.25 & 1794 \\ 
000653.31+000135.7 & 1.722100 &  0.02660 & 1.9481 & 5.5 & $-0.66$ & 5382 & 0.85 & 4554 \\
\hline
\end{tabular}
\footnotetext[1]{The full table is available online in a machine readable format.\label{fn1:full}}
\footnotetext[2]{Measured between 1710 and 3000~\AA\ ($f_\nu\propto\nu^{\alpha}$).\label{fn0:auv}}
\footnotetext[3]{The values of CF and FWHM are from the \HeII\ EW binning of the median non-BALQ emission. Binning by \auv\ yields nearly identical results (see text).\label{fn0:binning}}
\end{minipage}
\end{table*}

Tables~\ref{tab:HeII_bins} and \ref{tab:auv_bins} summarize the median properties of the \HeII\ EW and \auv\ bins, respectively. The tables also provide the median $L(3000\mbox{\AA})$, \MgII\ FWHM, \Mbh\ and \LLedd\ for each bin. The bins are similar in these properties. All bins have a roughly the same $L(3000\mbox{\AA}) \approx 10^{46}$~\ergs, and a similar $\Mbh \approx 10^9$~M$_{\odot}$ within a range of $\simeq\pm$0.1~dex. The median \LLedd\ is large ($>$0.3) in all bins, and spans a range of $\simeq$0.2~dex.

\begin{table*}
\begin{minipage}{155mm}
\caption{The median properties of the \HeII\ EW binned objects.}\label{tab:HeII_bins}
\begin{tabular}{@{}{l}*{11}{c}{c}@{}}
\hline
Class & $N_{\rm bin}$ & $N_{\rm obj}$ & \HeII &\auv\fnrepeat{fn1:auv}& $\log$ &   \MgII& $\log$ & $\log$ & $N_{\rm RL}$ &  \multicolumn{3}{c}{\CIV\ BAL\fnrepeat{fn1:EW}} \\
&  &  & EW & & $L(3000\mbox{\AA})$  &  FWHM& \Mbh & \LLedd  &  & CF & \vsp & FWHM  \\
& &  & (\AA) & & (\ergs)  & (\kms) & (M$_{\sun}$) & &   & & (\kms) & (\kms) \\
\hline
BALQs & 1 & 399 & 7.0 &$-0.97$ & 45.90 & 3900 & 9.03 & $-0.48$  & 15 & 0.73 & 5200 & 3000 \\
 & 2 & 399 & 4.7 & $-0.83$ & 45.97 & 4400 & 9.14 & $-0.52$  & 12 & 0.71 & 6200 & 3300 \\
 & 3 & 399 & 3.0 & $-0.79$ & 46.00 & 4100 & 9.09 & $-0.44$ & 19 & 0.75 & 7300 & 3700  \\
 & 4 & 399 & 0.9 & $-0.96$ &45.99 & 3400 & 8.93 & $-0.30$  &  43 & 0.80 & 8000 & 4300 \\
\hline
non- & 1 & 5330 & 7.3 & $-0.77$ & 45.81 & 3800 & 8.93 & $-0.49$ & 379 \\
BALQs & 2 & 3461 & 4.7 & $-0.65$ & 45.93 & 4000 & 9.03 & $-0.47$ & 146 \\
& 3 & 2425 & 3.1 & $-0.61$ & 45.94 & 3700 & 8.98 & $-0.38$ & 147\\
& 4 & 1368 & 1.0 & $-0.69$ &45.93 & 3300 & 8.87 & $-0.30$ & 108\\
\hline
\end{tabular}
\footnotetext[1]{Measured between 1710 and 3000~\AA\ ($f_\nu\propto\nu^{\alpha}$).\label{fn1:auv}}
\footnotetext[2]{The median value of individual BAL profiles (cf.\ BLH13, where the tabulated values are for the median BAL profile). \label{fn1:EW}}
\end{minipage}
\end{table*}

\begin{table*}
\begin{minipage}{155mm}
\caption{The median properties of the \auv\ binned objects.\fnrepeat{fn2:auv}}\label{tab:auv_bins}
\begin{tabular}{@{}{l}*{11}{c}{c}@{}}
\hline
Class & $N_{\rm bin}$ & $N_{\rm obj}$ & \auv&\HeII & $\log$ & \MgII & $\log$ & $\log$ & $N_{\rm RL}$ & \multicolumn{3}{c}{\CIV\ BAL\fnrepeat{fn2:EW}} \\
&  &  & &EW & $L(3000\mbox{\AA})$  &  FWHM& \Mbh & \LLedd &  & CF & \vsp & FWHM  \\
& &  & & (\AA) & (\ergs) & (\kms)  & (M$_{\sun}$) & &  & & (\kms) & (\kms)\\
\hline
BALQs & 1 & 399 & $-0.48$ &3.5 & 45.92 & 4100 & 9.08 & $-0.49$ & 9 & 0.69 & 8700 & 3500 \\
 & 2 & 399 & $-0.76$ & 4.0& 45.97 & 4100 & 9.09 & $-0.45$ & 17 & 0.69 & 7700 & 3600 \\
 & 3 & 399 & $-1.05$ &4.3 & 45.97 & 3900 & 9.05 & $-0.41$ & 15 & 0.76 & 5700 & 3500 \\
 & 4 & 399 & $-1.51$ &3.5 & 46.00 & 3600 & 9.02 & $-0.38$ & 48 & 0.82 & 4500 & 3300 \\
\hline
non- & 1 & 5414 & $-0.44$ &4.6 & 45.86 & 3800 & 8.96 & $-0.46$ & 249 \\
BALQs & 2 &  3334 & $-0.76$ &5.4 & 45.92 & 3800 & 8.99 & $-0.43$ & 174 \\
& 3 & 2349 & $-1.02$ &5.9 & 45.89 & 3700 & 8.97 & $-0.42$ & 171 \\
& 4 & 1526 & $-1.45$ &5.1 & 45.86 & 3600 & 8.92 & $-0.41$ & 181 \\
\hline
\end{tabular}
\footnotetext[1]{Measured between 1710 and 3000~\AA\ ($f_\nu\propto\nu^{\alpha}$).\label{fn2:auv}}
\footnotetext[2]{The median value of individual BAL profiles (cf.\ BLH13, where the tabulated values are for the median BAL profile). \label{fn2:EW}}
\end{minipage}
\end{table*}

Tables~\ref{tab:HeII_bins} and \ref{tab:auv_bins} also list the number of RL quasars ($N_{\rm RL}$) in each bin. The observed fraction of RL BALQs from the total RL quasar population increases with decreasing \HeII\ EW and a redder \auv. For the \HeII\ EW, the fraction increases from 4 per cent (15/394) to 28 per cent (43/151) from the highest to the lowest bin. For \auv, the fraction increases from 4 per cent (9/258) to 21 per cent (48/229) from the bluest to the reddest bin. The RQ population shows a similar increase, from 7 per cent (384/5335) to 22 per cent (356/1616) for the \HeII-EW binning, and from 7 per cent (390/5555) to 21 per cent (351/1696) for the \auv\ binning.

\subsection{Alternative HiBALQ samples}
We explore the effect of using only BALQs from the \citet{gibson_etal09} catalogue, where all BALQs are identified by an automated procedure, unlike the \citet{shen_etal11} catalogue (see above). There are 1287 HiBALQs in the \citet{gibson_etal09} catalogue which comply to our selection criteria.\footnote{The fraction of HiBALQs from the \citet{gibson_etal09} catalogue out of our whole HiBALQ sample is $1287/1596\simeq80$ per cent, which is similar to the fraction of BALQs from \citet{gibson_etal09} out of the whole BALQ sample of \citet{shen_etal11}.} This subsample of 1287 HiBALQs produces similar results to the whole sample of 1596 HiBALQs, both qualitatively and quantitatively.

We also check the effect of a subsample that includes only BALQs with a reported ${\rm BI}>0$ \citep{gibson_etal09}, which is a more `traditional' criterion of BALQ identification compared to ${\rm BI}_0>0$. The subsample contains 1046 HiBALQs, and produces similar trends to the whole HiBALQ sample. There are two main quantitative differences between the ${\rm BI}>0$ subsample and the whole sample. First, the ${\rm BI}>0$ subsample implies median values of \vsp\ which are $\sim$1000~\kms\ larger. This is expected since the ${\rm BI}>0$ criterion does not identify BALs in the $0<\vs<3000$~\kms\ range, in contrast with the ${\rm BI}_0>0$ criterion (see above). Second, a more marginal difference is that the ${\rm BI}>0$ subsample has median values of FWHM which are larger by 0--600~\kms\ than for the whole sample (the difference in the median value between the two samples with all bins combined is 200~\kms). Using the \citet{shen_etal11} sample, with BALQs detected by a visual inspection, allows us to examine trends across a wider range of outflow properties.

\section{Results}\label{sec:results}

\subsection{Distribution of the profile properties of \CIV\ BAL} \label{sec:results_ind}

\begin{figure*}
 \includegraphics[width=180mm]{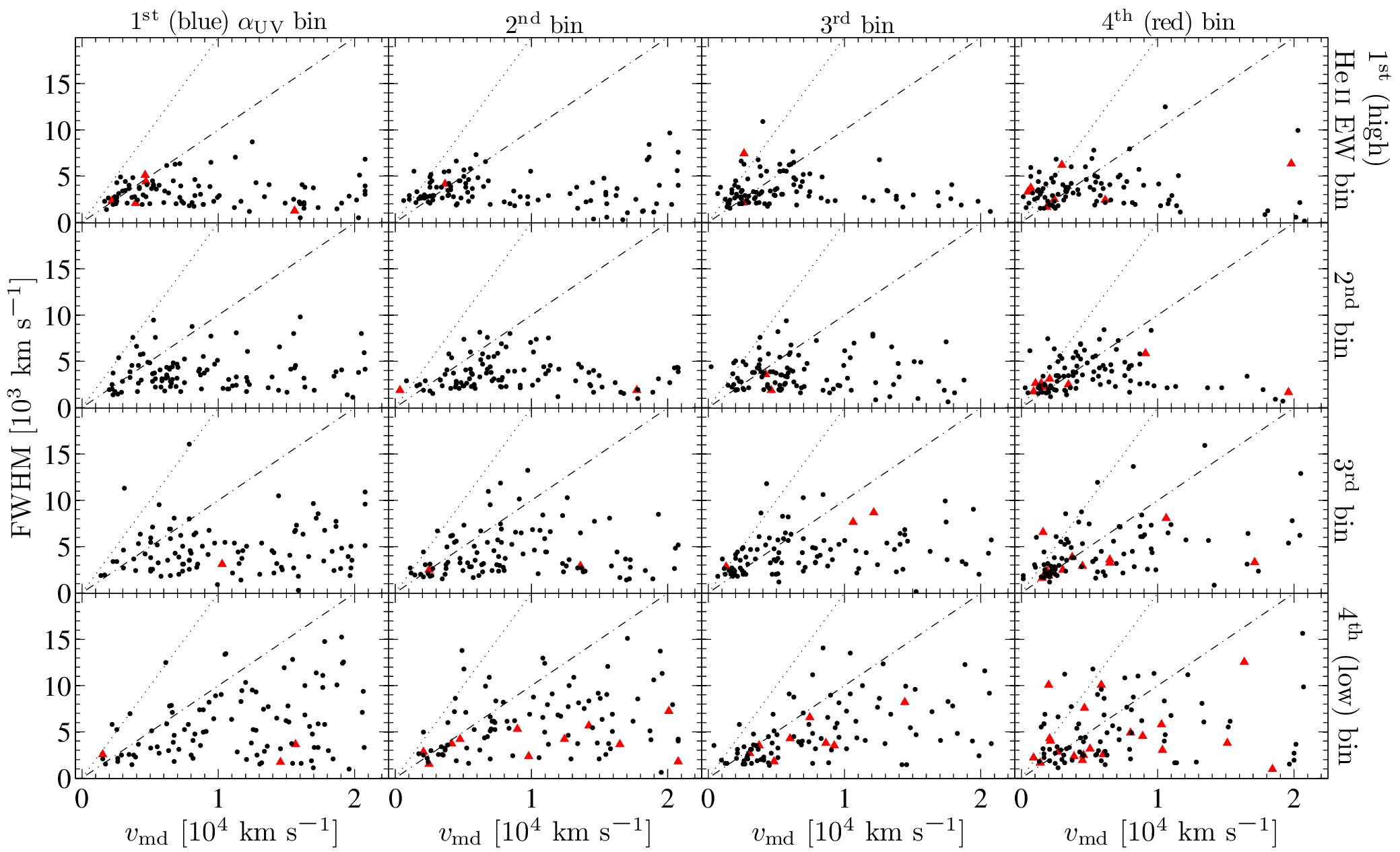}
\caption{
The distribution of FWHM versus \vsp\ of the \CIV\ BAL, for the \HeII\ EW and \auv\ bins. The BALQ and non-BALQ samples are first binned into \auv\ bins, and then each \auv\ bin is binned based on the \HeII\ EW. RL BALQs are denoted by a red triangle. The slope decreases (i.e.\ becomes redder) from left to right, and the \HeII\ EW decreases from top to bottom. The dot-dashed line and the dotted line denote 1:1 and 2:1 relation, respectively. Note the trend where in the reddest-\auv\ highest-\HeII\ EW bin, BALs cluster at the lowest values of \vsp\ and FWHM (most right-top panel), while in the bluest-\auv\ weakest-\HeII\ EW bin, BAL values are rather evenly spread (most left-bottom panel). 
}\label{fig:FWHM_vs_vshift}
\end{figure*}

Figure~\ref{fig:FWHM_vs_vshift} presents the distribution of FWHM versus \vsp\ of the \CIV\ BAL for the $4\times4$ \HeII\ EW and \auv\ bins. The BALQ and non-BALQ samples are first binned into four \auv\ bins, and then each \auv\ bin is binned into four bins based on the \HeII\ EW. For a given \HeII\ EW bin, \vsp\ reaches larger values as \auv\ becomes bluer, while the range of FWHM values remains approximately constant with \auv. For a given \auv\ bin, both \vsp\ and the FWHM reach larger values as the \HeII\ EW becomes weaker. The overall trend is that BALs cluster at the lowest values of \vsp\ and FWHM in the reddest-\auv\ highest-\HeII\ EW bin (top-right panel), while BAL values are rather evenly spread in the bluest-\auv\ weakest-\HeII\ EW bin (bottom-left panel). This trend is further explored below. For all bins, the FWHM is typically smaller than \vsp\ for $\vsp>4000$~\kms, except the $4\times 1$ reddest \auv\ bins (the rightmost panels in Fig.~\ref{fig:FWHM_vs_vshift}).

\begin{figure}
 \includegraphics[width=84mm]{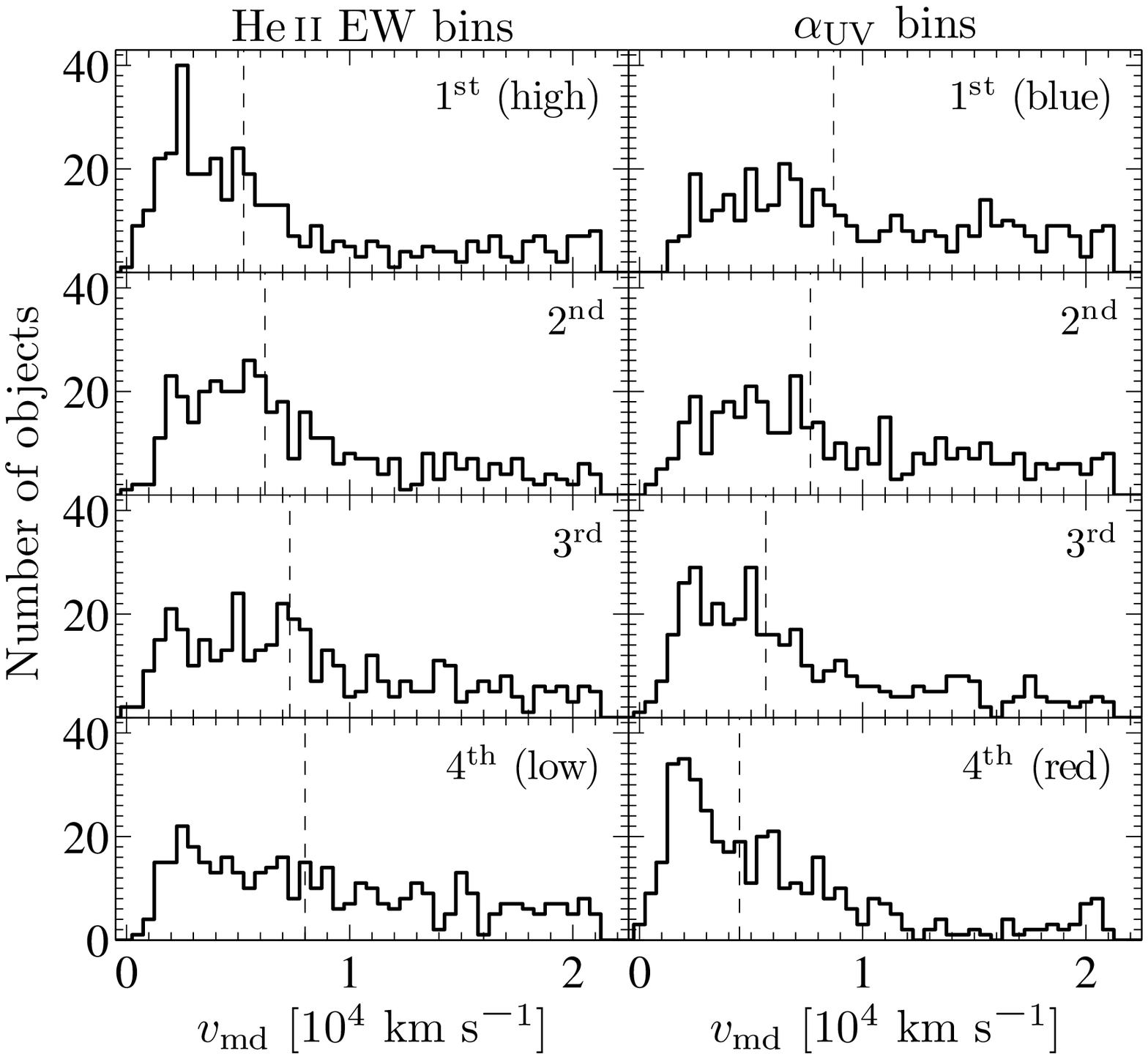}
\caption{
The distribution of \vsp\ of \CIV\ BAL in each \HeII\ EW and \auv\ bin. The median \vsp\ of the bin is denoted by the vertical dashed line. There are more \CIV\ BALs with high \vsp\ ($\ga10^4$~\kms) as the \HeII\ EW decreases (left panels), and as \auv\ becomes bluer (right panels).
}\label{fig:vshift_distr}
\end{figure}

Figure~\ref{fig:vshift_distr} presents a histogram of the distribution of \vsp\ values. The left panels present the distribution in each \HeII\ EW bin. In all bins, the distribution has a peak at $\vsp<5000$~\kms. As the \HeII\ EW becomes weaker, the distribution  becomes flatter, and the median value of \vsp\ increases from 5200~\kms\ for the highest \HeII\ EW bin to 8000~\kms\ for the lowest one (Table~\ref{tab:HeII_bins}). The right panels of Fig.~\ref{fig:vshift_distr} present the distribution of \vsp\ in each \auv\ bin. Here, the distribution becomes flatter as \auv\ becomes bluer. The median value of \vsp\ increases from 4500~\kms\ for the reddest bin to 8700~\kms\ for the bluest \auv\ bin (Table~\ref{tab:auv_bins}). The median value of \vsp\ for the whole BALQ sample is 6500~\kms.

\begin{figure}
 \includegraphics[width=84mm]{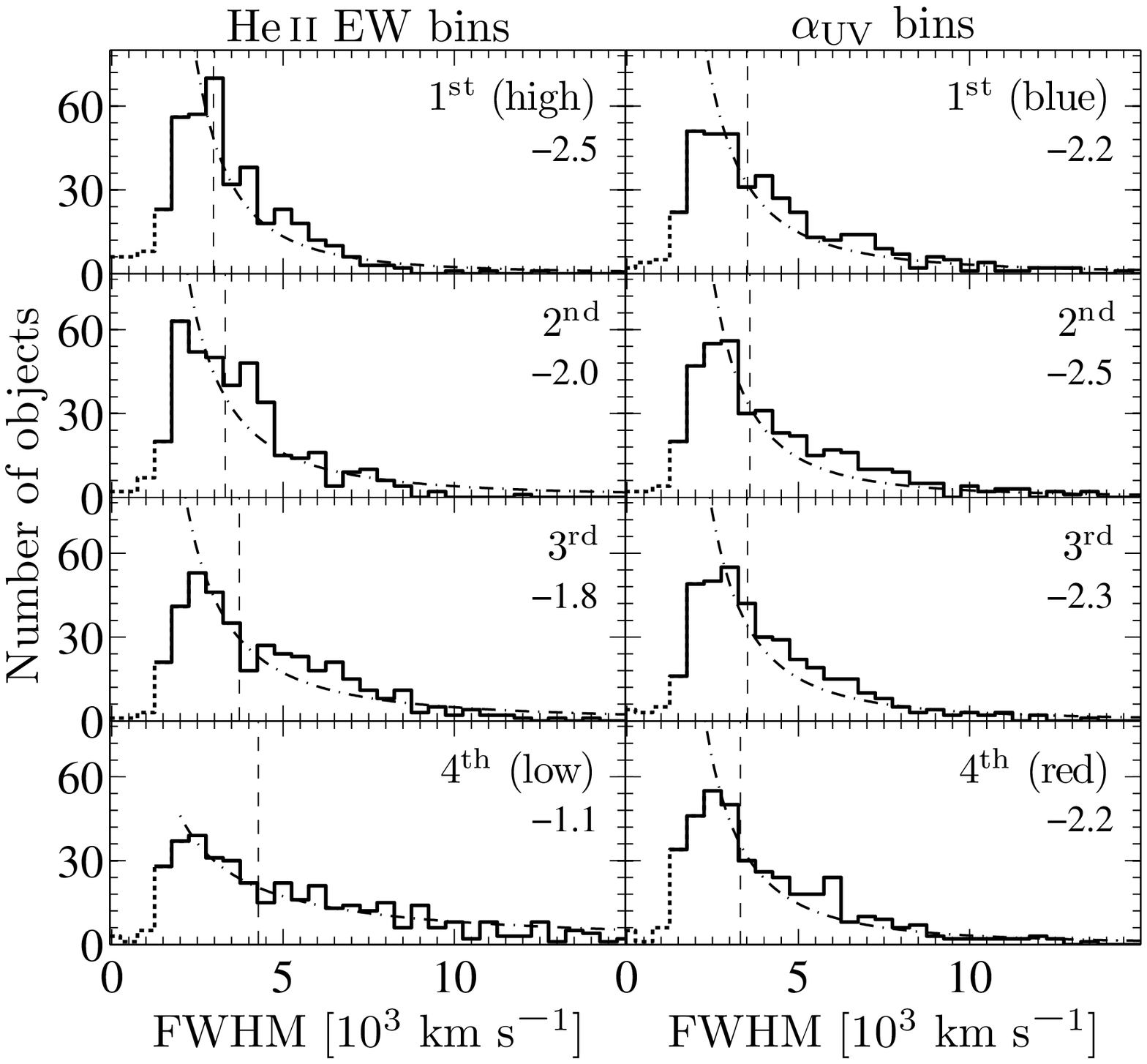}
\caption{
The same as Fig.~\ref{fig:vshift_distr}, for the FWHM of \CIV\ BAL. The dotted line marks the distribution for ${\rm FWHM}<1500$~\kms, i.e.\ smaller than the width of the smoothing filter that is applied to object spectra. The dot-dashed line is a power-law fit, which slope is denoted in the panel (see text). For all bins, the median ${\rm FWHM}\la 4000$~\kms, and the distribution falls off significantly by ${\rm FWHM}\la 10,000$~\kms. There are more \CIV\ BALs with ${\rm FWHM}>5000$~\kms\ as the \HeII\ EW decreases (left panels). The FWHM distribution is independent of \auv\ (right panels).
}\label{fig:FWHM_distr}
\end{figure}

Figure~\ref{fig:FWHM_distr} presents the distribution of FWHM values of the \CIV\ BAL. The left panels present the distribution in each \HeII\ EW bin. For all bins, the distribution decreases sharply with increasing FWHM for ${\rm FWHM}\ga 3000$~\kms, and the decrease is sharper for a higher \HeII\ EW. To quantify the decrease, we fit a power-law in the range FWHM = 2000--10,000~\kms\ for the two weaker bins, and between 2000--8000~\kms\ for the two stronger bins, which have a negligible number of objects with ${\rm FWHM}>8000$~\kms. The slope of the power-law decreases monotonically from $-1.1$ for the weakest \HeII\ EW bin, through $-1.8$ and $-2.0$, to $-2.5$ for the strongest one. The median value of FWHM increases from 3000~\kms\ for the strongest \HeII\ EW bin to 4300~\kms\ for the weakest bin (see Table~\ref{tab:HeII_bins}). The right panels of Fig.~\ref{fig:FWHM_distr} present the FWHM distribution in each \auv\ bin. The distribution is independent of \auv, as manifested by both the similar slopes  ($-2.2$ to $-2.5$; Fig.~\ref{fig:FWHM_distr}), and the similar median FWHM of $\simeq$3500~\kms\ for all \auv\ bins (Table~\ref{tab:auv_bins}). We also measure the velocity dispersion of the individual \CIV\ BAL absorption profiles, and its distributions show similar trends with the \HeII\ EW and \auv, as found above for the FWHM. The median FWHM of the whole BALQ sample is 3500~\kms. Only $\simeq$4 per cent of objects have a \CIV\ BAL with ${\rm FWHM}>10,000$~\kms, as seen in the BALQ PHL~5200 \citep*{junkkarinen_etal83}, which is often perceived as the `prototype' BALQ (e.g.\ \citealt{turnshek_etal88}).

\begin{figure}
 \includegraphics[width=84mm]{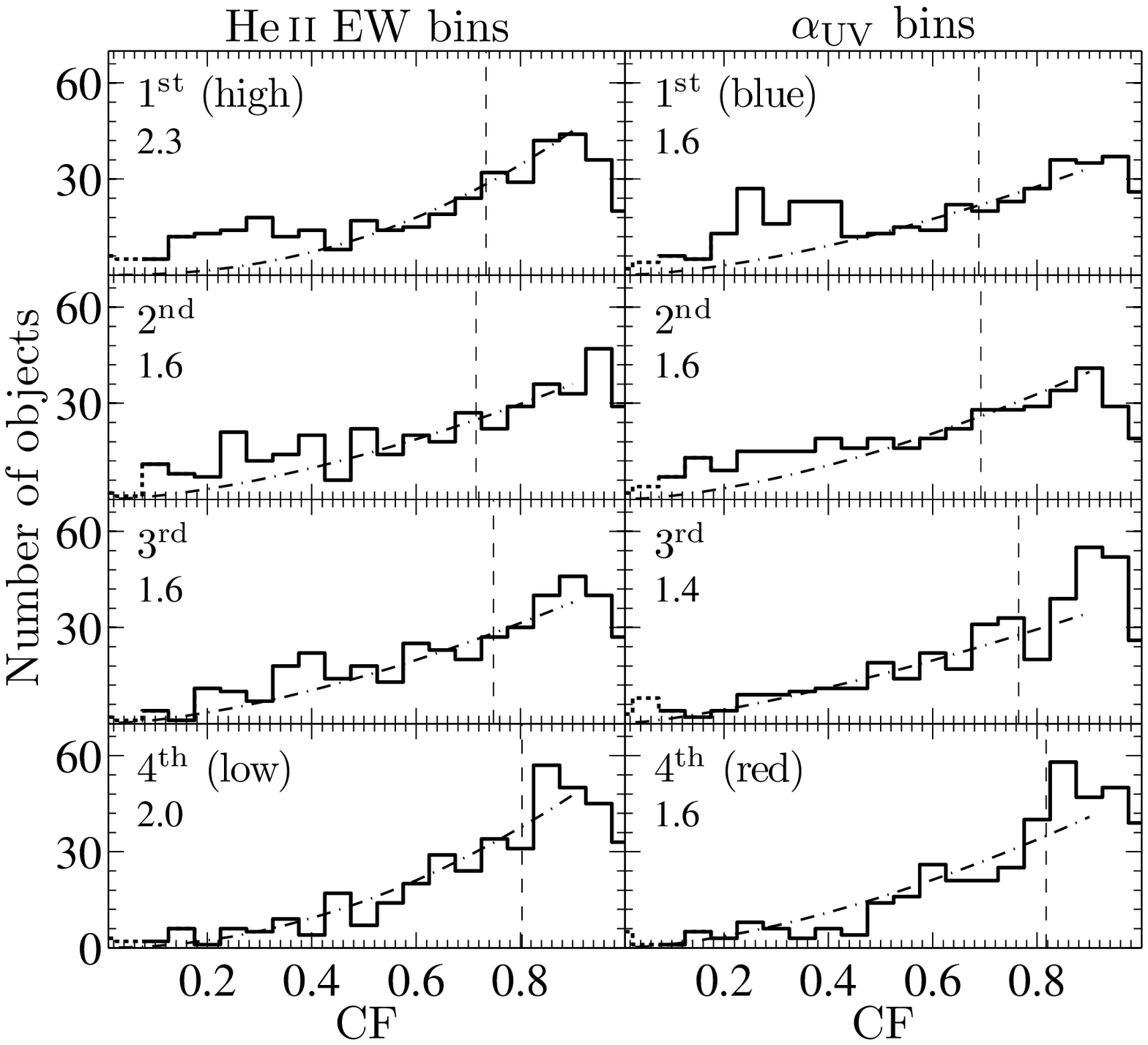}
\caption{
The same as Fig.~\ref{fig:FWHM_distr}, for the \CIV\ BAL CF. The dotted line marks the distribution for BAL CF $<$0.1, which implies ${\rm BI}=0$ (see Section~\ref{sec:data_analysis}). The power-law fit is between the CF of 0.5 and 0.8, and it has a similar slope for all bins. The excess of objects with a small value ($<0.5$) of CF relative to the power-law fit increases as the \HeII\ EW becomes larger, and as \auv\ becomes bluer. 
}\label{fig:distr_abs_at_vPeak}
\end{figure}

Figure~\ref{fig:distr_abs_at_vPeak} presents the distribution of the \CIV\ BAL CF. The CF distribution can be represented as being composed of two components. The first component, which is roughly similar for all bins, increases from ${\rm CF}=0$ as $\propto {\rm CF}^{1.6}$ (we fit a power-law in the range of CF = 0.5--0.8), and peaks at ${\rm CF}\approx0.9$. The second component is an excess of objects in the ${\rm CF} \approx 0.2-0.4$ range relative to the first component. This excess increases as the \HeII\ EW becomes stronger, and as \auv\ becomes bluer (Fig.~\ref{fig:distr_abs_at_vPeak}). The excess has a little effect on the median CF which increases slightly from $\simeq$0.7 to $\simeq$0.8 as \auv\ becomes redder and as the \HeII\ EW decreases. The excess has only a minor effect on the overall shape of the CF distribution for the \HeII\ EW binning, and the difference among the \HeII\ EW bins is statistically insignificant for most bin pairs (see below).

\begin{figure}
 \includegraphics[width=80mm]{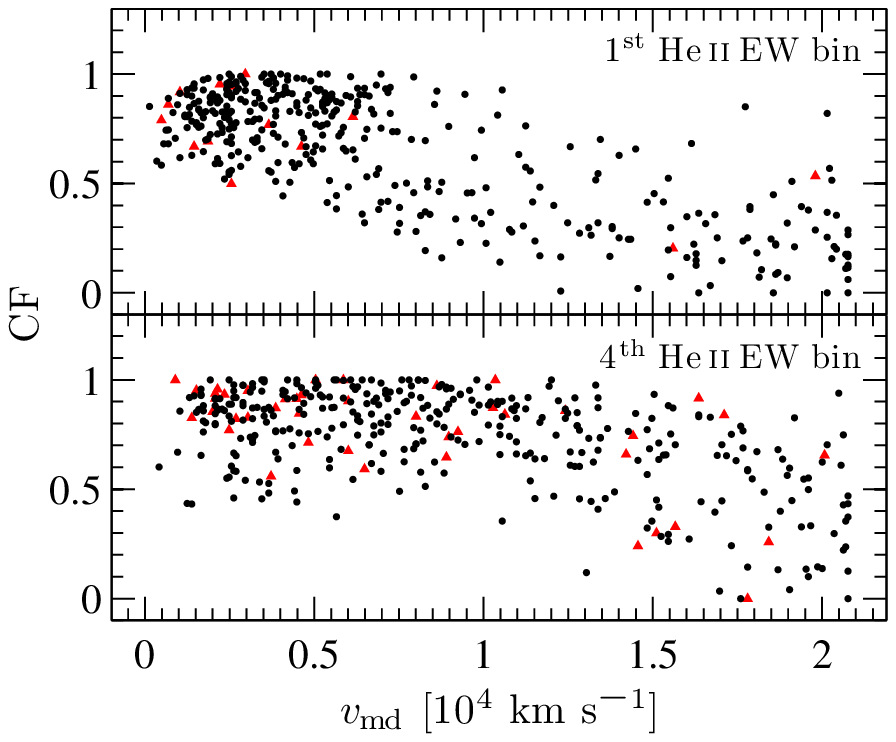}
\caption{
The \CIV\ BAL CF versus \vsp, for RQ (black dots) and RL (red triangles) BALQs. For all \HeII\ EW and \auv\ bins, the CF decreases as \vsp\ increases. RQ and RL BALQs show a similar distribution. We present only the highest and the lowest \HeII\ EW bins, for brevity. 
}\label{fig:abs_at_vPeak_vs_vPeak}
\end{figure}

Figure~\ref{fig:abs_at_vPeak_vs_vPeak} presents CF versus \vsp\ of the individual profiles for the highest and the lowest \HeII\ EW bins. The typical CF decreases as \vsp\ increases. A physical interpretation of this trend is presented in Section~\ref{sec:discussion}.

\subsubsection{Statistical tests}\label{sec:stat_tests}

Table~\ref{tab:rS_values} lists the values of \rS\ for correlations between the \CIV\ BAL properties and the \HeII\ EW and \auv. These values suggest that \vsp\ is more strongly controlled by \auv\ than by the \HeII\ EW ($\rS=0.30$ vs.\ $-0.15$); the FWHM depends on the \HeII\ EW, and is independent of \auv\ ($\rS=-0.23$ and $0.03$, respectively); and the CF has a slightly tighter correlation with \auv\ than with the \HeII\ EW, although both correlations are rather loose ($\rS=-0.17$ vs.\ $-0.13$).
Table~\ref{tab:rS_values} also lists the values of \rS\ for the relations between the different BAL properties. These values slightly depend on whether the BAL properties are derived using a non-BALQ composite matched in the \HeII\ EW or in \auv\ (Section~\ref{sec:data_analysis}). We list \rS\ for both matchings. There is a strong anti-correlation ($\rS\simeq-0.53$) between \vsp\ and CF, implying a decrease in CF with increasing \vsp\ (see also Fig.~\ref{fig:abs_at_vPeak_vs_vPeak}). The FWHM and CF have a positive correlation ($\rS\simeq0.34$). The correlation between \vsp\ and the FWHM is small ($\rS\simeq0.22$).

\begin{table}
\begin{minipage}{84mm}
\caption{The Spearman rank-order correlation coefficient (\rS) for the \CIV\ BAL properties.\fnrepeat{fn1:rS}$^,$\fnrepeat{fn2:rS}}\label{tab:rS_values}
\begin{tabular}{@{}{l}*{2}{c}{c}@{}}
\hline
 & \vsp\ &  FWHM & CF \\
 \hline
 \HeII\ EW & $-0.153$ & $-0.225$ & $-0.125$ \\
 \auv & 0.301 &  0.032 & $-0.171$ \\
 \vsp\ & & 0.216 & $-0.529$ \\
 FWHM & $0.158$ &  & 0.336 \\
 CF & $-0.548$ & $0.353$ & \\
\hline
\end{tabular}
\footnotetext[1]{The null probability is $p<10^{-6}$ for all relations, except the relation of FWHM with \auv\ for which $p=0.2$.\label{fn1:rS}}
\footnotetext[2]{The values of \rS\ for correlations among BAL properties slightly depend on whether the properties are derived using a non-BALQ composite matched in the \HeII\ EW or in \auv\ (Section~\ref{sec:data_analysis}). The values above the diagonal are for matching in the \HeII\ EW, and those below are for matching in \auv.\label{fn2:rS}}
\end{minipage}
\end{table}

We have applied the Kolmogorov-Smirnov (K-S) test to quantify the strength of the differences among distributions for the \HeII\ EW and \auv\ binning, for the three BAL parameters (Figs~\ref{fig:vshift_distr}--\ref{fig:distr_abs_at_vPeak}). For \vsp, the difference is statistically significant for most bin pairs ($p$-value $<$0.01; with $p<10^{-4}$ in most cases). The three exceptions are the first (bluest) \auv\ bin compared to the second one ($p=0.09$), and the third \HeII\ EW bin compared to the fourth (lowest) and the second one ($p=0.15$ and 0.02, respectively). For the FWHM, the difference is statistically significant for all \HeII\ EW bin pairs ($p<10^{-5}$ in most cases), except for the first bin compared to the second one ($p=0.15$). The difference is statistically insignificant for all \auv\ bin pairs ($p>0.55$). For the CF, the distribution for the \HeII\ EW binning is different with a high statistical significance only for the fourth bin compared to the other three ($p<10^{-3}$; $p>0.05$ for other bin pairs). For the \auv\ binning, the difference in CF distribution is statistically significant for all bin pairs ($p<10^{-2}$), except the first compared to the second bin ($p=0.45$).

The RL and RQ BALQ populations have a different distribution of the \HeII\ EW and \auv\ (K-S test yields $p\approx 10^{-6}$ and $10^{-9}$, respectively). RL BALQs have smaller values of \HeII\ EW and redder \auv. The median \HeII\ EW is 3.9 and 2.2~\AA\ for RQ and RL BALQs, respectively. The median \auv\ is $-0.88$ for RQ BALQs and $-1.26$ for RL BALQs. The distribution of the \CIV\ BAL FWHM and CF is similar for both populations ($p\simeq0.06$ and 0.02, respectively), while the distribution of \vsp\ is different ($p\approx10^{-4}$). Matching the RL and RQ BALQ populations in the distribution of both \HeII\ EW and \auv\ produces a similar distribution for both population in all three \CIV\ BAL parameters. The K-S test results in $p\simeq 0.23$, 0.01 and 0.46 for \vsp, FWHM and CF distribution, respectively, implying that these parameters are likely drawn from the same distributions for both RL and RQ BALQs.

\subsection{Median BAL profiles -- the quasar rest-frame versus the \CIV\ BAL rest-frame}

\begin{figure*}
 \includegraphics[width=120mm]{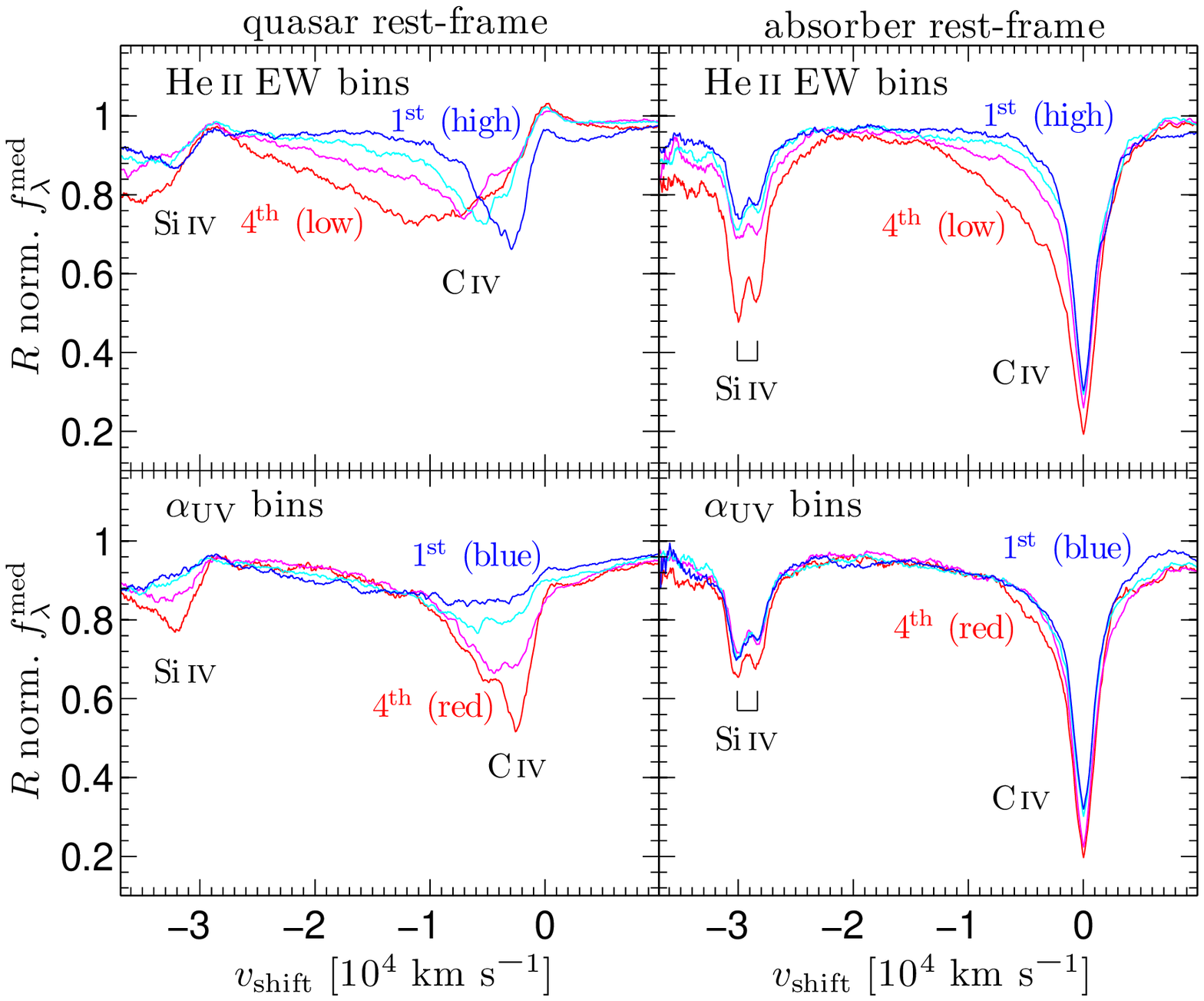}
\caption{
Comparison between the median \CIV\ and \SiIV\ BAL profiles in the quasar (left panels; an adaptation of fig.~9 of BLH13) and in the \CIV\ absorber (right panels) rest-frame. The BAL profiles are presented for four bins of \HeII\ EW and for four bins of \auv\ (top and bottom pales, respectively). The maximum depth of the \CIV\ BAL profile in the absorber rest-frame is at $\vs=0$ by construction. For composites in the quasar rest-frame, the \CIV\ BAL profile becomes broader and shifts to a higher \vs, as the \HeII\ EW becomes lower; and the BAL profile becomes deeper as \auv\ becomes redder. The FWHM of the median profile increases by $\simeq$12,000~\kms\ from the highest to the lowest \HeII\ EW bin. The median BAL profile is broad (${\rm FWHM}\geq5000$~\kms) for all \HeII\ EW and \auv\ bins. In contrast, for composites in the absorber rest-frame, the \CIV\ BAL profile is narrow (${\rm FWHM}\leq4000$~\kms) for all bins; the FWHM of the median profile increases only by $\simeq$1000~\kms\ from the highest to the lowest \HeII\ EW bin; and \auv\ has only a small effect ($\simeq$15 per cent) on the BAL depth. These differences are expected from the difference in the distribution of \vsp\ and FWHM with the \HeII\ EW and \auv\ (Figs~\ref{fig:vshift_distr} and \ref{fig:FWHM_distr}). The \SiIV\ BAL absorption that is associated with the \CIV\ absorber increases as the ionizing SED becomes softer (i.e.\ a smaller \HeII\ EW; upper-right panel), as expected for a photoionized gas. 
}\label{fig:zoom_on_CIV} 
\end{figure*}

Figure~\ref{fig:zoom_on_CIV} presents the median \CIV\ and \SiIV\ BAL profiles for the \HeII\ EW and \auv\ bins. For a given bin, the median profile is evaluated by dividing the BALQ composite by the corresponding non-BALQ composite. The median profile is evaluated for two types of rest-frame. The first rest-frame is the quasar rest-frame (left panels), i.e.\ both BALQ and non-BALQ spectra are aligned based on $z$ prior to calculating the composite (as done in BLH13). The second rest-frame is the rest-frame of the \CIV\ BAL absorber (right panels), i.e.\ the BALQ spectra are aligned based on \vsp, which is used to define $\vs=0$ (see Section~\ref{sec:data_analysis}). When calculating the non-BALQ composite, the non-BALQ spectra are shifted by \vs\ that has the same distribution as \vsp. 

Fig.~\ref{fig:zoom_on_CIV} presents a dramatic difference in the median \CIV\ BAL profile between the two types of rest-frame. For the quasar rest-frame, the median BAL profile becomes broader by $\simeq$12,000~\kms\ as the \HeII\ EW becomes weaker; and the absorption becomes deeper as \auv\ becomes redder (as further discussed in BLH13). In contrast, for the \CIV\ absorber rest-frame, the BAL profile remains narrow (${\rm FWHM}<4000$~\kms) for all bins; the FWHM of the median profile increases by only $\simeq$1000~\kms, from $\simeq$2500 to 3500~\kms, as the \HeII\ EW decreases; and \auv\ has only a small effect ($\simeq$15 per cent) on the BAL CF. The CF increases from 0.7 to 0.8, both as \auv\ becomes redder and as the \HeII\ EW decreases. The FWHM remains $\simeq$3000~\kms\ for all \auv\ bins. The large difference in the median profiles between the two rest-frames results from the dependence of the \vsp\ distribution on the \HeII\ EW and \auv\ (Fig.~\ref{fig:vshift_distr}). In the bluer \auv\ bins, \vsp\ extends to higher velocities, compared to the redder bins. As a result, the individual \CIV\ BAL profiles are spread over a larger range of velocities, yielding a shallower and broader median profile, although the individual profiles are nearly identical in the \vsp\ frame.  A similar effect takes place in the \HeII\ EW based bins.

It is interesting to note that the \SiIV\ BAL absorption strength (seen at $\vs\approx -30,000$~\kms) increases significantly as the \HeII\ EW becomes weaker (Fig.~\ref{fig:zoom_on_CIV}, upper-right panel), with the median CF increasing from $\sim$0.2 to $\sim$0.5. The \SiIV\ doublet lines have a different absorption strength, indicating that the \SiIV\ absorption is not fully saturated.

\subsection{Comparison with previous studies}
There are only a few studies which explore the distribution of properties of the \CIV\ BAL profile in a large sample of BALQs. For a sample of BALQs  from the SDSS DR5, \citet{gibson_etal09} report that the total width of \CIV\ BAL is typically smaller than 5000~\kms\ (fig.~9 there). This is consistent with the median FWHM being smaller than 4000~\kms\ for all \HeII\ EW and \auv\ bins (Fig.~\ref{fig:FWHM_distr}). \citet{gibson_etal09} also find that the distribution of \vs\ reaches its maximum below $\vs=8000$~\kms, and then gradually decreases with increasing \vs. These findings are in accordance with the distribution of \vsp\ (Fig.~\ref{fig:vshift_distr}). \citet{allen_etal11} explore the distribution of the mean absorption depth of \CIV\ BAL for BALQs from the SDSS DR6. They find that the distribution peaks at $\simeq$0.6, which is roughly consistent with the maximum absorption of $\simeq 0.7-0.8$, implied by the composite profile in the absorber rest-frame (Fig.~\ref{fig:zoom_on_CIV}, right panels).

\section{Discussion}\label{sec:discussion}

\emph{Why does the \CIV\ absorption profile depend on the \HeII\ EW?} As noted above (Section~\ref{sec:intro}), the \HeII\ emission EW is likely an indicator of the hardness of the ionizing SED, since it measures the continuum strength above 54~eV relative to the near-UV continuum at $\lamrest=1640$~\AA. A higher value of \HeII\ EW corresponds to a harder SED, and vice-versa. This interpretation naturally explains the observed increase of \SiIV\ BAL strength with decreasing \HeII\ EW (Fig.~\ref{fig:zoom_on_CIV}, upper-right panel). For a line-driven outflow, a softer SED implies a larger force multiplier for a given ionization state, and thus produces more favourable conditions for wind launching. A trend between the outflow $v$ and the \HeII\ EW is implied by the mass flux continuity equation, i.e.\ $nv={\rm const.}$, where an outflow at a larger $v$ has a lower density $n$. For a low enough $n$, the outflow becomes overionized and ceases to produce \CIV\ absorption. For a softer ionizing SED (i.e.\ a weaker \HeII\ EW), overionzation occurs at a lower $n$, and thus at a higher $v$.\footnote{The implied rise of the ionization parameter with $v$ may also be produced by another mechanism \citep{fukumura_etal10a, fukumura_etal10b}.} If the outflow is line driven, then the larger force multiplier for a softer SED results in a higher outflow velocity.

The interpretation of the \HeII\ EW as an indicator of the SED hardness can readily explain the observed trends. The interpretation implies that objects with a harder SED (a larger \HeII\ EW) will produce outflows with a smaller \vsp\ and a narrower \CIV\ BAL profile, as indeed observed (Figs~\ref{fig:FWHM_vs_vshift}--\ref{fig:FWHM_distr}). The BAL profile is narrower for a harder SED, since the gas that produces the absorption wings is already overionized at a higher $n$, and thus at a lower $v$. The interpretation is also consistent with the trend of the distribution of BAL CF with the \HeII\ EW (Fig.~\ref{fig:distr_abs_at_vPeak}, left panels). The distribution has a tail of smaller values of BAL CF which increases as the \HeII\ EW becomes larger. The tail may represent a BALQ population with a non-saturated \CIV\ absorption ($\tau\sim1$), which is expected to be larger for a harder SED that can overionize larger columns of gas. Alternatively, the observed CF may be produced by the projected area of inhomogeneous gas clumps with $\tau\gg1$ \citep{hamann_etal11}. This area decreases for a harder SED, as the outer parts of the clumps become overionized, producing the observed tail of small values of CF.

\emph{Why does the \CIV\ absorption profile depend on \auv?} The value of \auv\ is likely an indicator of our viewing angle, with a bluer \auv\ indicating an angle closer to face-on. This interpretation assumes that \auv\ is mostly set by reddening, and that the dust preferably resides in the symmetry plane of the quasar, i.e.\ the plane of the accretion disc (AD; see BLH13, section 5.2, for a detailed justification of this interpretation). The trend of the distribution of \vsp\ with \auv, where there are more objects with a high \vsp\ for a bluer \auv\ (Fig.~\ref{fig:vshift_distr}), may also be a viewing angle effect. This relationship between the acceleration and the viewing angle $\theta$ relative to the normal to the AD is expected, for example, for a line-driven outflow, since the driving UV continuum flux is expected to go as $\cos\theta$ for AD emission, and thus to produce a larger accelerating force for a smaller $\theta$ (see also \citealt{proga_03, kurosawa_proga08}).

The lack of a trend between the distribution of the \CIV\ BAL FWHM and \auv\ (Fig.~\ref{fig:FWHM_distr}), implies that the mechanism that produces the BAL velocity dispersion, as indicated by the absorption FWHM, is mostly isotropic and is independent of the bulk motion direction. It was recently reported by \citet{hamann_etal13} for mini-BALs with extreme values of \vsp\ that the absorbing gas may be in a form of a thin `pancake'-like filament. If this form also applies to regular BALs, as suggested by theoretical considerations \citep*{baskin_etal14b}, then the BAL velocity dispersion probably originates from ordered motions which are internal to the filament and resemble turbulence.

The distribution of CF becomes bimodal for bluer \auv\ slopes, and the bluest bin has two peaks, at ${\rm CF}\simeq0.9$ and 0.3 (Fig.~\ref{fig:distr_abs_at_vPeak}). The peak at ${\rm CF}\simeq0.3$ may result from a larger projected area of the emitting region for the blue (face-on) objects, which makes it harder for the outflow to cover the whole emitting region, compared to objects viewed edge-on. The typical BAL CF of $\simeq$0.7--0.8 at maximum absorption (Fig.~\ref{fig:zoom_on_CIV}) may result from ${\rm CF} \approx 1$ that is `diluted' by scattered continuum from a medium with ${\rm CF}\simeq0.2-0.3$, rather than from partial covering. A conclusive evidence for this scenario is provided by observations that find a high polarization level of the light at the bottom of the absorption troughs \citep*{glenn_etal94, cohen_etal95, goodrich_miller95, ogle_97, schmidt_hines99, ogle_etal99, lamy_hutsemekers00, lamy_hutsemekers04, brotherton_etal06}. The ${\rm CF}\simeq0.2-0.3$ required for the scattering medium is similar to the measured CF of 0.3 for the Broad Line Region (BLR; \citealt{korista_etal97b, maiolino_elal01, ruff_etal12}). Thus, it is likely that the BLR is the scattering medium \citep{goodrich_miller95, korista_ferland98}, if the BAL outflow resides at a distance comparable to that of the BLR. The presence of significant structure within the BAL profile clearly demonstrates that it is not completely dominated by scattered light, which produces a trough with no structure.

The mass flux continuity likely explains the decrease in BAL CF with increasing \vsp, which is observed for all bins of \HeII\ EW and \auv\ (Fig.~\ref{fig:abs_at_vPeak_vs_vPeak}). The ionization level of the absorber increases with \vsp, since the absorber $n$ decreases with increasing \vsp, from continuity. The increase of ionization level results in an overall drop in $\tau$. Either $\tau<1$ in a uniform absorber, or the CF of the part of the clumpy outflow with $\tau>1$ becomes smaller.

The trends of the \CIV\ BAL profile parameters with the \HeII\ EW and \auv, which we find for HiBALQs, favour the scenario in which there is a smooth transition from HiBALQs to LoBALQs (e.g.\ \citealt{elvis00}). LoBALQs are observed to have a redder continuum than HiBALQs \citep{weymann_etal91, boroson_meyers92, sprayberry_foltz92, reichard_etal03, gibson_etal09}, and are likely predominantly located at the parameter-space of the lowest \HeII\ EW and the reddest \auv\ (BLH13). The increase in the CF of \CIV\ BAL with decreasing \HeII\ EW and redder \auv\ (Fig.~\ref{fig:distr_abs_at_vPeak}) implies that the \CIV\ absorption in LoBALQs should be typically stronger than in HiBALQs, as indeed observed \citep{reichard_etal03, allen_etal11, filizAk_etal14}. \citet{allen_etal11} report that LoBALQs have a broader \CIV\ BAL than HiBALQs. This is consistent with the increase of \CIV\ BAL FWHM with decreasing \HeII\ EW (Fig.~\ref{fig:FWHM_distr}). The increase in the strength of the intermediate-ionization \SiIV\ BAL with decreasing \HeII\ EW (Fig.~\ref{fig:zoom_on_CIV}) also favours the scenario of smooth transition between HiBALQs and LoBALQs.

\subsection{Comparison between RL and RQ BALQs}
\emph{Do RL BALQs differ in the \CIV\ BAL properties from RQ BALQs?} It has been suggested that RL BALQs have different BAL properties compared to RQ BALQs (e.g.\ \citealt{becker_etal97,brotherton_etal98}; cf.\ \citealt{rochais_etal14}). We find that this difference does not result from radio loudness, but rather from the preference of RL BALQs to have a weak \HeII\ EW and a red \auv\ (Fig.~\ref{fig:binning}, and Tables~\ref{tab:HeII_bins} and \ref{tab:auv_bins}; see also \citealt{rochais_etal14}, fig.~6 there). When RL and RQ BALQs are matched in the \HeII\ EW and \auv, both subclasses show similar \CIV\ BAL properties (Section~\ref{sec:stat_tests}; see also Figs~\ref{fig:FWHM_vs_vshift} and \ref{fig:abs_at_vPeak_vs_vPeak}). \citet{richards_etal11} reach a similar conclusion for the broad emission lines. They match RL and RQ non-BALQs by luminosity, $z$ and the \CIV\ emission EW and blueshift, and find no significant difference in other emission lines between the two subclasses. As we show below, the \CIV\ emission EW and blueshift have a strong correlation with the \HeII\ EW (see also Baskin, in preparation).

RL non-BALQs have a higher \HeII\ EW and a bluer \auv\ compared to RL BALQs, similarly to the RQ subclass (Tables~\ref{tab:HeII_bins} and \ref{tab:auv_bins}). However, RL non-BALQs are redder than RQ non-BALQs. The median \auv\ is $-0.84$ for RL non-BLAQs and $-0.69$ for RQ non-BALQs; and 23 per cent of RL non-BALQs reside in the reddest bin, compared to only 11 per cent of RQ non-BALQs. In contrast, the median \HeII\ EW is similar for both RL and RQ non-BALQ populations (5.4 and 5.1~\AA, respectively), and both populations are similarly distributed among the \HeII\ EW bins.

The observed BALQ fraction from the total RL quasar population increases from 4 (4) per cent in the highest \HeII\ EW (bluest \auv) bin to 28 (21) per cent in the lowest \HeII\ EW (reddest \auv) bin. The complete quasar sample, which is dominated by RQ quasars ($\simeq$94 per cent of BALQs and non-BALQs), produces a comparable observed BALQ fraction in the corresponding bins (BLH13; see also Section~\ref{sec:data_analysis}). This further supports the suggestion that radio-loudness does not directly affect BAL properties.

The preference of RL BALQs to lie in the parameter-space of weak \HeII\ EW and red \auv\ explains the large fraction of LoBALQs in RL BALQ samples, as LoBALQs have preference for the same parameter-space (see above). \citet{becker_etla00} find a LoBALQ fraction of $\simeq$0.5 in the FIRST BALQ sample, compared to $\sim$0.1 found in SDSS samples that are dominated by RQ BALQs. \citet{dipompeo_etal11} report that RL BALQs are typically viewed at larger angles relative to the radio-jet axis compared to RL non-BALQs. This supports the interpretation of \auv\ as an inclination angle indicator, where a redder \auv\ implies a larger angle, since the radio-jet axis is likely to be perpendicular to the symmetry plane of the system.

\subsection{Similar relations for the Broad Line Region \CIV\ emission}

There are observational indications that the BLR emission originates in two components, a disc and a wind component \citep{Collin_Souffrin_etal88, wills_etal93, leighly_moore04, richards_etal11, kruczek_etal11}. Recently, \citet{baskin_etal14b} noted that if the BAL gas is located at distances similar to those of the BLR, then it likely contributes to the emission of broad lines. This BAL gas may very well be the optically-thin BLR gas that was suggested by \citet*{shields_etal95}. Does the wind component of the BLR show similar trends to those of BAL outflows?

\begin{figure}
 \includegraphics[width=80mm]{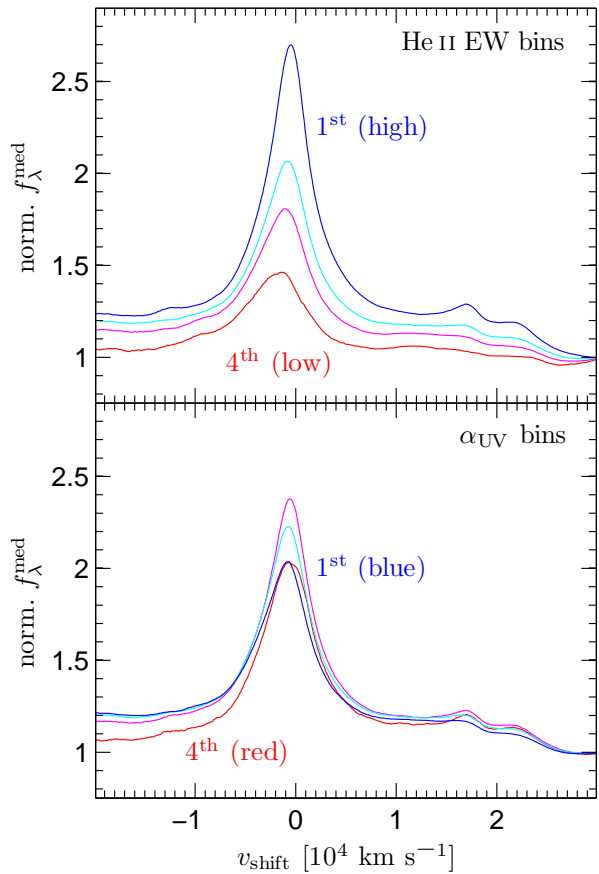}
\caption{
The dependence of the \CIV\ emission-line profile of \emph{non-BALQs} on the \HeII\ EW and \auv\ (top and bottom panel, respectively). The \HeII\ EW affects both the \CIV\ strength and shape. As the \HeII\ EW becomes weaker, the \CIV\ emission becomes weaker, more blue-shifted and more asymmetric. For the first three \auv\ bins, the \CIV\ emission becomes stronger as \auv\ becomes redder.
}\label{fig:zoom_on_CIV_emission} 
\end{figure}

Figure~\ref{fig:zoom_on_CIV_emission} presents the \CIV\ emission-line profile for the \emph{non-BALQ} bins of \HeII\ EW and \auv. The same non-BALQ bins from the BAL analysis are adopted (i.e.\ the non-BALQ sample is not rebinned), and the plotted \CIV\ profiles correspond to the unabsorbed \CIV\ emission of matching BALQ bins. The \CIV\ emission becomes weaker and more blueshifted and asymmetric as the \HeII\ EW becomes weaker (Fig.~\ref{fig:zoom_on_CIV_emission}, top panel). This trend results from the transition of \CIV\ emission from being dominated by the disc component (high \HeII\ EW objects) to being dominated by the wind component (low \HeII\ EW objects). A similar trend is presented in \citet[fig.~11 there]{richards_etal11}. The total \CIV\ emission decreases with a decreasing \HeII\ EW, since the dominant emitter of \CIV\ is the disc component, which is likely radiation bounded. As the SED becomes softer (lower \HeII\ EW), the production of C$^{3+}$ decreases for a radiation-bounded gas, and the \CIV\ emission EW becomes lower (e.g.\ \citealt*{baskin_etal14a}). Radiation-bounded gas absorbs all of the ionizing radiation, and thus the line emission should be stronger for a harder SED that has more ionizing photons, as observed (Fig.~\ref{fig:zoom_on_CIV_emission}, top panel). A similar trend of decreasing \CIV\ emission with a decreasing \HeII\ EW is observed for the red wing of \CIV\ in BALQs (BLH13). The \CIV\ emission from the wind component is marginally stronger for a softer SED (lower \HeII\ EW; Baskin, in preparation), since this component is likely matter bounded. As the SED becomes softer, a matter-bounded wind ($\tau<1$) produces C$^{3+}$ down to a lower $n$, and yields \CIV\ emission that extends up to a higher \vs\ (from mass flux continuity), as observed. The relation between the total \CIV\ line emission from the wind and the SED depends on the wind density structure, and requires detailed calculations. A possible connection between the BLR wind component and the BAL outflow will be addressed in a future study (Baskin, in preparation).

The trend of \CIV\ emission with \auv\ is consistent with \auv\ being an indicator of our viewing angle. The \CIV\ EW becomes larger for a redder \auv\ for the first three bins, and then decreases for the fourth reddest bin (Fig.~\ref{fig:zoom_on_CIV_emission}, bottom panel; note that all composites are normalized to 1 at $\lamrest=1700$~\AA). The BLR emission is likely only weakly dependent on inclination $\theta$. In contrast, the underlying UV continuum from the AD goes as $\cos\theta$. The combined effect results in a decrease of the continuum flux relative to the \CIV\ flux, i.e.\ an increase of the \CIV\ EW, as the inclination angle gets closer to edge-on (i.e.\ a redder \auv). The trend of increasing \CIV\ EW with a redder \auv\ is consistent with \citet{richards_etal02}.

Our findings (Fig.~\ref{fig:zoom_on_CIV_emission}) are consistent with the correlations reported by \citet{baskin_laor05} for a complete and well defined sample of Palomar-Green quasars \citep{boroson_green92}. The reported correlation between \aox\ and the \CIV\ EW (see also \citealt{kruczek_etal11}) is consistent with \CIV\ becoming weaker for a lower \HeII\ EW. The anti-correlation between the optical-UV slope and the \CIV\ EW is consistent with \CIV\ becoming stronger with a redder \auv. The anti-correlation between the \CIV\ blueshift and its EW (see also \citealt{richards_etal02, richards_etal11}) is detected here for both the \HeII\ EW and \auv\ binning. We leave the full analysis of the non-BALQ sample to a separate study (Baskin, in preparation).

The range of \HeII\ EW values for objects in our sample (Fig.~\ref{fig:binning}) can be produced by the observed range of ionizing SED slopes (i.e.\ $\alpha_{\rm ion}$ from $-2.0$ to $-1.2$, where $\alpha_{\rm ion}$ is defined between $\lamrest=912$ and 12~\AA; \citealt{baskin_etal14a}, fig.~5 there). A relation between the \HeII\ EW and the SED may be indicated by the strong relation of \HeII\ EW with the quasar luminosity $L$ \citep{boroson_green92, dietrich_etal02}, and the relation between $L$ and the SED (\citealt{scott_etal04, strateva_etal05, steffen_etal06, just_etal07, stern_laor12}; cf.\ \citealt{telfer_etal02, shull_etal12}). \citet*{korista_etal97a} suggested that the BLR does not see the same SED we do. They adopt a baseline of 0.1 for the BLR CF, and use the Locally Optimally emitting Cloud (LOC) model of the BLR, where a single slab of gas produces line emission from a rather limited range of ionization states \citep{baldwin_etal95}. However, adopting the observed BLR CF of $\simeq$0.3 (see above), and noting that a single slab produces line emission from a very broad range of ionization states \citep{baskin_etal14a}, alleviate the need in a different SED for the BLR. The interpretation of \HeII\ EW as an indicator of SED hardness can be directly tested by observations that cover the extreme-UV range (e.g.\ \citealt{stevans_etal14}).

\section{Conclusions}\label{sec:conclusions}

We analyse the distribution of the \CIV\ BAL profiles for a sample of BALQs from the SDSS DR7, which covers the wavelength range of 1400--3000~\AA. In contrast with our earlier study (BLH13), where we explored the median absorption profile in the quasar emission frame, here we present the median absorption profile in the outflow frame, defined by the velocity at maximum depth of absorption. The median absorption profile in the absorber frame is quite different from the median in the quasar emission frame.  We find the following:
\begin{enumerate}

 \item Although the typical wind velocity is $\vsp\simeq6500$~\kms, the velocity dispersion is small, with a median ${\rm FWHM}=3500$~\kms. Only 4 per cent of BALQs have ${\rm FWHM}>10,000$~\kms. 
 
 \item The \HeII\ emission-line EW affects the distribution of the wind velocity dispersions.  The distribution peaks at ${\rm FWHM}<3000$~\kms, and  decreases approximately as a power law at ${\rm FWHM}>3000$~\kms. The power law slope is  flat ($-1.1$) when the \HeII\ EW is low ($\simeq1$~\AA), and steep ($-2.5$) when the ${\rm EW}$ is high ($ \simeq7$~\AA). In contrast, the FWHM distribution is not affected by the \auv\ slope.
 
 \item The wind outflow velocity extends to higher values of \vsp\ as the \HeII\ EW decreases, from a median value of \vsp\ of 5200~\kms\ for ${\rm EW}\simeq7$~\AA, to a median value of 8000~\kms\ for ${\rm EW}\simeq1$~\AA. The outflow velocity extends to higher values also for a bluer \auv, from 4500~\kms\ for $\auv\simeq-1.5$, to 8700~\kms\ for $\auv\simeq-0.5$.
 
 \item The distribution of the wind CF increases from ${\rm CF}=0$ as $\propto {\rm CF}^{1.6}$, and peaks at ${\rm CF}=0.9$. The distribution does not depend on either the \HeII\ EW or \auv. There is an additional component which contributes only at the ${\rm CF}<0.5$ range, and its contribution is larger for a higher \HeII\ EW and a bluer \auv.
 
 \item Radio-loud BALQs have similar outflow properties to RQ BALQs when the two subclasses are matched in the \HeII\ EW and \auv. The reported difference in BAL properties between RL and RQ BALQs results from the preference of RL BALQs to have a smaller \HeII\ EW and a redder \auv\ compared to RQ BALQs.
 
 \item The \CIV\ emission line of \emph{non-BALQs} also shows trends with the \HeII\ EW and \auv. The symmetric \CIV\ emission, which is attributed to the `disc component' of the BLR, decreases with decreasing \HeII\ EW. The line becomes more asymmetric and blueshifted as the \HeII\ EW decreases. The line EW increases with increasing reddening, while its velocity profile is roughly independent of \auv.

\end{enumerate}

 The trends with the \HeII\ EW described above are consistent with the interpretation that the \HeII\ EW is an indicator of the hardness of the ionizing SED (extreme-UV relative to far-UV). For a softer SED (a lower \HeII\ EW), the BAL outflow extends to a lower $n$, and thus to a higher $v$ (from mass flux continuity), producing \CIV\ BAL profiles that are broader, and have a higher \vs\ and a larger CF, as observed.  The trends with \auv\ are consistent with the interpretation that \auv\ is an indicator of our viewing angle $\theta$. For a viewing angle closer to face-on (a bluer \auv), the AD continuum flux that goes as $\cos\theta$ is larger, and a line-driven wind can be accelerated to larger velocities, as observed. For this viewing angle, the emitting region has a larger projected area, which makes it harder for the outflow to fully cover the emitting region, and produces BALs with ${\rm CF}<0.5$. The mechanism that sets the BAL FWHM is likely independent of the viewing angle.
 
As pointed out in BLH13, the interpretation of \auv\ as an indicator of viewing angle can be tested by searching for a relationship between \auv\ and the continuum polarization, in both non-BALQs and BALQs. This interpretation also implies a trend between \auv\ and the X-ray absorbing column, which can be looked for. Spectra from the Cosmic Origins Spectrograph on board the Hubble Space Telescope, which cover the wavelength range of $\simeq$500--2000~\AA\ \citep{stevans_etal14}, can be utilized to directly test the interpretation of the \HeII\ EW as an indicator of the SED hardness.

\section*{Acknowledgements}
We thank the anonymous referee and G.\ Richards for valuable comments and suggestions. This research was supported by the Israel Science Foundation (grant No.~1561/13). This research has made use of the Sloan Digital Sky Survey which is managed by the Astrophysical Research Consortium for the Participating Institutions; and of NASA's Astrophysics Data System Bibliographic Services. 

Funding for the SDSS and SDSS-II has been provided by the Alfred P.\ Sloan Foundation, the Participating Institutions, the National Science Foundation, the U.S.\ Department of Energy, the National Aeronautics and Space Administration, the Japanese Monbukagakusho, the Max Planck Society, and the Higher Education Funding Council for England. The SDSS Web Site is http://www.sdss.org/.

The SDSS is managed by the Astrophysical Research Consortium for the Participating Institutions. The Participating Institutions are the American Museum of Natural History, Astrophysical Institute Potsdam, University of Basel, University of Cambridge, Case Western Reserve University, University of Chicago, Drexel University, Fermilab, the Institute for Advanced Study, the Japan Participation Group, Johns Hopkins University, the Joint Institute for Nuclear Astrophysics, the Kavli Institute for Particle Astrophysics and Cosmology, the Korean Scientist Group, the Chinese Academy of Sciences (LAMOST), Los Alamos National Laboratory, the Max-Planck-Institute for Astronomy (MPIA), the Max-Planck-Institute for Astrophysics (MPA), New Mexico State University, Ohio State University, University of Pittsburgh, University of Portsmouth, Princeton University, the United States Naval Observatory, and the University of Washington.

\bsp
\label{lastpage}
\end{document}